\documentclass[12pt,preprint]{aastex}

\shorttitle{Formation of multiple stellar populations}
\shortauthors{Jiang, Han \& Li}

\begin{document}

\title{Binary interactions as a possible scenario for the formation of multiple stellar populations in globular clusters}

\author{Dengkai Jiang\altaffilmark{1,2}, Zhanwen Han\altaffilmark{1,2} and Lifang Li\altaffilmark{1,2}}

\altaffiltext{1}{Yunnan Observatories, Chinese Academy of Sciences,
Kunming, 650011, China; dengkai@ynao.ac.cn, zhanwenhan@ynao.ac.cn}

\altaffiltext{2}{Key Laboratory for the Structure and Evolution of
Celestial Objects, Chinese Academy of Sciences}

\begin{abstract}

Observations revealed the presence of multiple stellar populations
in globular clusters (GCs) that exhibit wide abundance variations
and multiple sequences in Hertzsprung-Russell (HR) diagram. We
present a scenario for the formation of multiple stellar populations
in GCs. In this scenario, initial GCs are single-generation
clusters, and our model predicts that the abundance anomalous stars
observed in GCs are the merged stars and the accretor stars produced
by binary interactions, which are rapidly rotating stars at the
moment of their formation and are more massive than normal single
stars in the same evolutionary stage. We find that due to their own
evolution, these rapidly rotating stars have different surface
abundances, effective temperatures and luminosities from normal
single stars in the same evolutionary stage. The stellar population
with binaries can reproduce two important observational evidences of
multiple stellar populations, the Na-O anticorrelation and the
multiple sequences in HR diagram. This suggests that binary
interactions may be a possible scenario for the formation of
multiple stellar populations in GCs.

\end{abstract}


\section{Introduction}
Observational evidences for multiple stellar populations in globular
clusters (GCs) challenge the traditional view of GCs hosting simple
stellar populations, which are formed in one generation from a cloud
of uniform composition (single-generation cluster). One important
evidence is the presence of star-to-star abundance variations. The
most famous example is the Na-O anticorrelation that oxygen-depleted
stars have higher sodium abundances, which is exhibited among the
main sequence (MS) turnoff (TO) stars, sub-giant branch (SGB) stars
and red-giant branch (RGB) stars in some GCs
\citep{Carretta2004,Gratton2004}. Another important evidence is that
some GCs show multiple MSs \citep{Piotto2007}, multiple SGBs
\citep{Milone2008,Marino2009}, or anomalously wide RGB
\citep{Yong2008,Lee2009} in the Hertzsprung-Russell (HR) diagram.

Several scenarios have been proposed to explain the formation of
multiple stellar populations \citep{Gratton2012}, and they can be
classified into two contrasting hypotheses: \emph{Primordial}
Scenario vs. \emph{Evolutionary} Scenario
\citep{Kraft1994,Carretta2004}. In the \emph{Primordial} Scenario,
multiple stellar populations in GCs are formed in the different
materials that are primordial chemical inhomogeneities, for example,
a merger of two separate GCs \citep{Lee1999,Mackey2007} or the
self-pollution of the intra-cluster gas occurring in the early
evolution of clusters
\citep{D'Antona2002,Decressin2007,deMink2009a}. However, a merger
event would not be expected to occur in the halo of the Milky Way
due to the very large relative velocities of GCs
\citep{Gratton2012}. For the the self-enrichment of GCs, it is
difficult to explain the high fraction of the second-generation
formed from the gas polluted by the first-generation stars
\citep{Bastian2009}, and the nature of possible polluters remains
unclear \citep{Sills2010,Gratton2012}.

In the \emph{Evolutionary} Scenario, initial GCs are believed to be
single-generation clusters, and multiple stellar populations in GCs
are attributed to the evolution of stars affected by some physical
processes, such as the first dredge-up, Sweigart-Mengel Mixing
\citep{Sweigart1979}, primordial rotation \citep{Bastian2009}.
However, it is necessary to explain why only some of stars in GCs
are affected by these physical processes. Moreover,
\citet{Decressin2007} suggested that the MS TO stars in GCs are not
hot enough for producing the observed Na-O anti-correlation, because
the central temperature (about 25 $\times$ $10^6$\,K) of a
0.85\,M$_{\rm \odot}$ TO stars is higher than the required
temperature of CNO cycle (about 20 $\times$ $10^6$\,K), but is lower
than that of the production of $^{23}$Na from proton-capture on
$^{20}$Ne (about 35 $\times$ $10^6$\,K).

Binary interactions have been used to explain the formation of
multiple stellar populations as a kind of \emph{Primordial}
Scenario. Chemically enriched material ejected from massive binaries
have been proposed to form a second population of stars
\citep{deMink2009a}, or to be accreted by low-mass pre-main-sequence
stars \citep{Bastian2013}. However, binary interactions are not
considered as a kind of \emph{Evolutionary} Scenario. They are not
investigated to produce the abundance anomalous stars
\emph{directly}, although they have been proposed as an explanation
for extended MS in the HR diagrams of intermediate age clusters
\citep{Yang2011,Li2012}.

Binary systems can produce stars more massive than normal single
stars in the same evolutionary stage by binary interactions, e.g.
the merged stars or the accretor stars. These stars are not hot
enough for the production of $^{23}$Na from proton-capture on
$^{20}$Ne, but they are hot enough for proton-captures on $^{16}$O
and $^{22}$Ne, which respectively lead to the destruction of
$^{16}$O and to the production of $^{23}$Na at temperatures $>$20
$\times$ $10^6$\,K \citep{Prantzos2006}. Furthermore, these stars
produced by binary interactions will be rapid rotating because
binary interactions can convert orbital angular momentum into their
spin angular momentum \citep{deMink2013}. Rotationally induced
mixing can bring part of the nuclear-processed products from the
core to the surface \citep{Decressin2007}, and stellar rotation was
used to explain the observed nitrogen enrichment found in several
massive MS stars \citep{Brott2009}. Therefore, these rapidly
rotating stars reproduce the properties of the abundance anomalous
stars observed in GCs (e.g. the Na-O anticorrelation), and binary
interactions will be a possible scenario for the formation of
multiple stellar populations in GCs as a kind of \emph{Evolutionary}
Scenario.

In this paper, we adopt a simple model for this scenario to
investigate the formation of multiple stellar populations. In
Section 2, we explore the formation of rapidly rotating stars from
binary interactions by performing binary evolution calculations, and
calculated the subsequent evolution of these rapidly rotating stars
to obtain their parameters (e.g. effective temperatures,
luminosities and surface abundances). Then, we estimate the
distributions of the binary population by performing the binary
population synthesis study. In Section 3, we give discussion and
conclusions.

\section{THE MODEL AND SIMULATION}
In our scenario, the anomalous stars observed in GCs are the merged
stars and the accretor stars produced by binary interactions, and
these binary products are rapidly rotating and more massive than the
normal stars. These binary products have different effective
temperatures, luminosities and surface abundances from single stars
in the same evolutionary stage due to their own evolution. Thus, the
stellar population with binaries (the binary population) can
reproduce the observations of multiple stellar populations, because
the binary population includes two populations: (1) normal single
stars with normal abundance (initial abundance of GCs); (2) the
merged stars and the accretor stars (produced by binary
interactions) that have abnormal surface abundances by their own
evolution.

\subsection{Binary evolution calculations}

We considered three evolutionary channels for the formation of
rapidly rotating stars from binary interactions: (1) a binary
evolves into contact and then merges into a rapidly rotating star
(contact merger channel for the merged stars)
\citep{Jiang2012,deMink2013}; (2) a binary evolves into Roche lobe
overflow (RLOF) and experiences unstable mass transfer while both
components are still MS stars, and then merges into a rapidly
rotating star (unstable RLOF merger channel for the merged stars)
\citep{Jiang2012}; (3) a binary evolves into RLOF and experiences
stable mass transfer, and the secondary obtains the mass and angular
momentum transferred from the primary, and then becomes a rapidly
rotating star (stable RLOF channel for the accretor stars)
\citep{deMink2013}.

In these channels, a binary needs to evolve into contact phase or
unstable RLOF phase while both components are still MS stars, or
evolve into stable RLOF phase while the secondary is still a MS star
\citep{Jiang2012,deMink2013}. This is because for MS stars, the
binding energy of the envelope might be large enough to form rapidly
rotating stars. To determine whether the binary evolves into these
phase, it is necessary to perform detailed binary evolution
calculations. We use Eggleton's stellar evolution code
\citep{Eggleton1971,Eggleton1972,Eggleton1973} that has been updated
with the latest input physics during the last four decades
\citep{Han1994,Pols1995,Pols1998,Nelson2001,Eggleton2002}. In our
calculations, we used the metallicity $Z = 0.001$ (corresponding to
typical GCs) and considered the nonconservative effects, includes a
model of dynamo-driven mass loss, magnetic braking and tidal
friction for the evolution of stars with cool convective envelopes
\citep{Eggleton2002}.

Altogether, we have calculated the evolution of 5200 binary systems,
thus obtained a large, dense model grid that covers the following
ranges of initial primary mass ($M_{10}$), initial mass ratio ($q_0
= M_{20}/M_{10}$) and initial orbital period ($P_0$):
log\,$M_{10}=-0.3, -0.25, . . . , 0.2$;
log\,$q_0=$log\,$(M_{20}/M_{10})=0.05, 0.10, . . . , 0.5$; and
log\,$(P_0/P_{\rm ZAMS})= 0.05, 0.10, . . . $, where $P_{\rm ZAMS}$
is the orbital period at which the initially more massive component
would just fill its Roche lobe on the zero-age MS. These ranges of
initial parameters cover the systems that can form rapidly rotating
stars in old stellar population by Case A binary evolution.
According to these binary models, the parameters of rapidly rotating
stars are established when the binary evolves into contact, unstable
RLOF, or the end of stable RLOF. For simplicity, in the contact
merger channel and the unstable RLOF merger channel, we assume that
there are 10\% mass loss during the merger phase
\citep{Jiang2012,deMink2013} and the binary products rotate at 80\%
of break-up velocity. For the stable RLOF channel, we calculate the
angular momentum obtained by the secondary, and then calculate its
rotational velocity. We only consider the secondaries that can get
enough angular momentum to rotate at 80\% of break-up velocity.

\subsection{Rotation evolution calculations}
To calculate the subsequent evolution of rapidly rotating stars
produced by binary interactions, we have used the Modules for
Stellar Experiments in Astrophysics code (MESA version 4631)
\citep{Paxton2011}. This code implements the effects of rotation in
the transport of angular momentum and chemical mixing, and it has
been used to study the effects of rotational mixing
\citep{Chatzopoulos2012a,Chatzopoulos2012b}. We do not consider the
effect of the helium-rich core of the original components on the
subsequent evolution of rapidly rotating stars, and we assume that
these rapidly rotating stars have abundances similar to zero age MS
stars. In addition, we consider that the effect of rotation on
low-mass stars with convective envelopes might be weaker due to
magnetic braking \citep{Hurley2000,Bastian2009}. We calculate the
stellar evolutionary models with $M = 0.6, 0.7 ... 1.6$\,M$_{\rm
\odot}$ for $Z = 0.001$ that cover rapidly rotating stars produced
by binary interactions in old stellar populations. For simplicity,
these models run for two different degrees of rotation: non-rotating
and 80\% of break-up velocity, which are used to investigate the
evolution of surface abundances of single stars and rapidly rotating
stars produced by binary interactions. All evolutionary tracks have
been computed up to the point of core helium ignition.

To illustrate the evolutionary tracks of these stars, we show four
evolution examples of surface mass fractions (sodium versus oxygen
abundances) in Fig. 1. During the evolution of rapidly rotating star
with $M=1.3$\,M$_{\rm \odot}$ (dotted line), the surface mass
fraction of sodium ($X_{\rm Na23}$) increases from $1.22\times
10^{-6}$ to $5.95\times 10^{-6}$ , and that of oxygen ($X_{\rm
O16}$) decreases from $4.67\times 10^{-4}$ to $9.54\times 10^{-5}$.
This is because the destruction of O and the production of $^{23}$Na
in the central region result from proton-captures on $^{16}$O and
$^{22}$Ne \citep{Langer1993,Prantzos2006,Decressin2007}, and they
are brought to the surface by rotational mixing. In other examples,
rapidly rotating stars with $M=1.0, 1.1, 1.2$\,M$_{\rm \odot}$
evolve in a similar way as in the previous example, although the
variation ranges of surface abundances decrease with decreasing
mass. Hence, these rapidly rotating stars can be oxygen depletion
and sodium enhancement. For single stars with no rotation (open
circles in Fig. 1), the surface mass fractions of sodium and oxygen
do not show significant variations.

\subsection{Binary population synthesis}
In the binary population synthesis study, we have performed a Monte
Carlo simulation. We adopt the following input for the simulation
(see Han et al. 1995). (1) The initial mass function (IMF) of
\citet{Miller1979} is adopted. (2) The mass-ratio distribution is
taken to be constant. (3) We take the distribution of separations to
be constant in log\,$a$ for wide binaries, where $a$ is the orbital
separation. Our adopted distribution implies that $\sim$50 per cent
of stellar systems are binary systems with orbital periods less than
100\,yr. We follow the evolution of a million sample binaries
according to grids of binary models and the evolution models of
rapidly rotating stars described above.

The simulation gives the distributions of many properties of the
binary population, e.g. the masses, the effective temperatures, the
luminosities, the surface abundances, etc. Fig. 2 are selected
distributions of the binary population at 10 Gyr that may be helpful
for understanding the scenario of binary interactions for the
formation of multiple stellar populations. Fig. 2(a) shows the
distribution of single stars and binary products in the Na-O plane.
Single stars (open circles) are concentrated at the bottom right
corner and do not show significant variations. However, binary
products have large star-to-star variations. The distribution of
binary products first extends upward to high sodium abundance region
from the original abundance location, and then turns left to the
oxygen depletion region. Thus, the binary population with binary
interactions can show a Na-O anticorrelation.

To investigate the distributions of binary products with different
abundances in HR diagram, we classify them into four groups based on
the surface mass fraction of Na and O, which are denoted as pluses
with different colours. The distribution of the binary population
with binary interactions in HR diagram is shown in Fig. 2(b). It is
obvious that this population has at least two MSs, three TOs, three
SGBs and extended RGB, which are produced by single stars (open
circles) and binary products (pluses). More importantly, these
sequences have a large spread in sodium and oxygen abundances. This
is consistent with the fact that the Na-O anticorrelation is
observed among MS TO stars, SGB stars and RGB stars in some GCs
\citep{Carretta2004,Gratton2004}. Observed red giants in 17 GCs
given by \citet{Carretta2009} are shown in Fig. 2(c), and they have
a distribution similar to that of our simulated binary populations.
If we assume that the binary populations (black line and blue lines)
with different initial abundances, such as in different GCs, have
the same variation ranges of surface abundances, then we would find
that these binary populations in the simulation are in good
agreement with the observed stars in 17 GCs as shown in Fig. 2(c).

We noted that single stars and binary products in the binary
population have different mass distributions as shown in Fig. 2(d),
and binary products are more massive than single stars. For the
binary population with log $L/L_{\rm \odot}
> 0$, the mass range of single stars is from 0.75 to 0.9\,M$_{\rm \odot}$,
while the mass range of binary products is from 0.8\,M$_{\rm \odot}$
to 1.35\,M$_{\rm \odot}$, although these single stars and binary
products have the same range of evolutionary stages. The maximum
mass of binary products is larger than that of single stars by
$\sim0.45$\,M$_{\rm \odot}$. We do not compare stars with log
$L/L_{\rm \odot} < 0$ because the sample of very low-mass binary
products is not complete in our calculations. Although the effect of
rotation on low-mass stars might be weaker due to magnetic braking
\citep{Hurley2000,Bastian2009}, massive binary products could have
significantly different surface abundances from single stars. Binary
products are more massive than single stars in the same evolutionary
stage, and they rotate rapidly, i.e. close to break-up velocity. The
destruction of $^{16}$O and the production of $^{23}$Na in the
central region result from proton-captures on $^{16}$O and
$^{22}$Ne, and the products of nuclear burning are brought to the
surface by rotational mixing. This can explain why the abundance
variations take place in these stars, and not in single stars at the
same evolutionary stage.

\section{Discussion and conclusions}
In this paper, we presented binary interactions as a possible
scenario for the formation of multiple stellar populations in GCs.
In this scenario, binary interactions can convert orbital angular
momentum into spin angular momentum and produce rapidly rotating
stars. These stars might have different properties from single
stars, such as surface abundances, temperatures and luminosities,
because they experienced binary interactions and their evolution is
affected by rotationally induced mixing. The existence of binary
products and single stars results in multiple stellar populations in
GCs, although initial GCs are single-generation clusters. In our
simple model, we carried out binary evolutionary calculations and
constructed the evolutionary models of rapidly rotating stars formed
by binary interactions. By performing a Monte Carlo simulation, we
obtained the distributions of the binary population and investigated
the formation of multiple stellar populations from binary
interactions.

Stellar rotation have been shown that can explain the broad MS and
spread TO observed in intermediate age stellar clusters
\citep{Bastian2009, Li2012}. However, \citet{Bastian2009} suggested
that this is unlikely to cause multiple stellar populations observed
in old GCs, because the stars in GCs are low mass and are not
expected to be rapidly rotating stars. We show that old binary
population could show multiple sequences in HR diagram as shown in
Fig. 2(b), because binary interactions could produce rapidly
rotating stars. Binary interactions, including mass transfer
\citep{Pols1991,deMink2013} or merger \citep{Tylenda2011,Jiang2013},
have been used to investigate the formation of rapidly rotating
stars. Moreover, observed evidences have shown that binary
interactions are very common in GCs. Many contact binaries have been
observed in GCs \citep{Rucinski2000}, and these binaries are
believed to merge into a rapidly rotating star (contact merger
channel). \citet{Mathieu2009} show that many of blue stragglers are
rotating faster than normal MS stars , which might be formed by mass
transfer \citep{Geller2011}. Therefore, binary interactions could
explain multiple sequences in HR diagram, which is important
evidence of multiple stellar populations.

Other important evidence of multiple stellar populations in GCs is
the presence of star-to-star abundance variations, and the most
famous example is the Na-O anticorrelation. We show that old binary
population can reproduce a Na-O anticorrelation as a result of the
formation of rapidly rotating stars from binary interactions. If the
binary populations with different initial abundances, such as in
different GCs, are assumed to have the same variation ranges of
surface abundances, we found that these binary populations in the
simulation are in good agreement with the observed stars in 17 GCs
given by \citet{Carretta2009}. In our scenario, this anticorrelation
could be found in TO stars, SGB stars and RGB stars, which is in
agreement with the observation that show the Na-O anticorrelation in
these evolutionary stages \citep{Carretta2004,Gratton2004}.
Moreover, our model predicts that these abundance anomalous stars
should be more massive than MS TO stars, and they should be blue
stragglers or blue straggler progeny. Therefore, binary interactions
could explain the existence of the Na-O anticorrelation in GCs. This
suggests that binary interactions may be a possible scenario for the
formation of multiple stellar populations, and the traditional view
of single-generation cluster could explain the photometric and
spectroscopic observations of multiple stellar populations by
considering the effects of binary interactions.

Observed multiple stellar populations show the cluster-to-cluster
variations \citep[e.g.][]{Carretta2009}, and these variations also
need to be explained \citep{Sills2010}. This cluster-to-cluster
variations might be due to the dependence of binary interactions on
the cluster properties, such as age, initial abundance, initial
fraction of binaries and initial orbital separation distribution.
These parameters determine the nature of binary interactions and the
subsequent evolution of binary products. When the initial fraction
of binaries is very low, the binary population would have few binary
products and be similar to a simple, single stellar population. The
investigation of the properties of binaries in 13 low-density GCs
revealed that the fraction of binaries ranges from 0.1 to 0.5 in the
core depending on the cluster \citep{Sollima2007}. Moreover, the
stellar density in GCs is sufficiently high to affect the binaries
that can be destroyed, created or modified by dynamical interaction
\citep[e.g.][]{Hut1992}. Higher stellar densities could result in
more efficient binary dissolution and thus a lower binary fraction
\citep{Marks2011}. Consequently, the binary population might have
different distributions and different fractions of binary products
in various clusters. A more detailed study of the binary population
with binary interactions might help to understand the dependence of
binary interactions on the cluster properties in the further
studies.

It is clear that our model is still quite simple and does not
investigate in much detail so far. In our investigation, we do not
consider the effect of the helium-rich core of the original
components on the subsequent evolution of binary products. Our
simulation shows that 36 per cent of all rapidly rotating stars are
formed by binary merger (including contact merger channel and
unstable RLOF merger channel). These rapidly rotating stars formed
by binary merger might contain the helium-rich core of the original
primary, and their remaining relative lifetimes are smaller than
those of single stars with the same masses
\citep[e.g.][]{Sills1997,Glebbeek2008,deMink2013}. For rapidly
rotating stars produced through stable RLOF, this effect need not to
be considered because they are the less-massive components of
binaries and their cores are still hydrogen rich at the onset of
mass transfer \citep[e.g.][]{Sills1997,Glebbeek2008,deMink2013}. For
rotating velocity, we simply assume that all binary products rotate
initially at 80\% of break-up velocity. A lower initial rotating
velocity or a faster spin-down will lead to a weaker variation of
surface abundances. These effects need more detailed investigation
in further study. In addition, the observations suggested that the
fraction of binaries in GCs may be lower than that in the field and
open clusters \citep{Rubenstein1997}. This might be because GCs are
much older than the field and open clusters, and consequently more
binaries in GCs are not expected to be recognized as binaries due to
their large luminosity ratio and their large orbital periods, or
they may not even be binaries due to merger \citep{deMink2013}.

\acknowledgments

We thank an anonymous referee for his/her valuable comments. This
work was supported by the Natural Science Foundation of China (Nos
11073049, 11033008, 11103073, 11390374 and 11373063), Science and
Technology Innovation Talent Programme of the Yunnnan Province (No.
2013HA005) and the Western Light Youth Project.

\appendix

\clearpage
\begin{figure}
\epsscale{0.6} \plotone{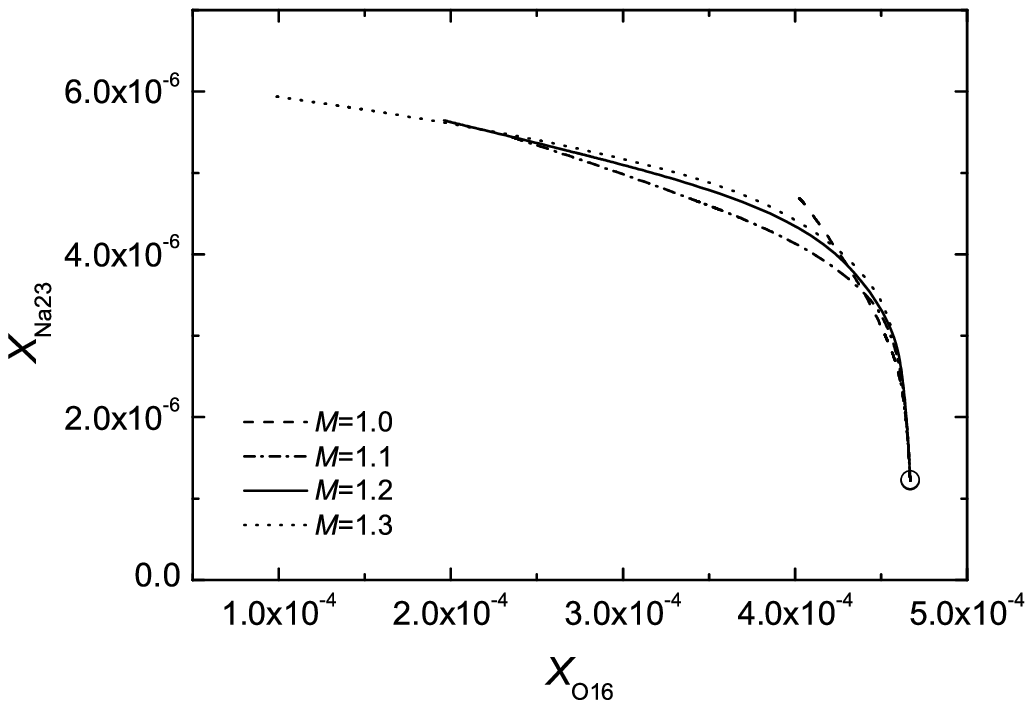}\caption[]{Evolutionary examples of
rapidly rotating stars from their formation in the surface mass
fraction ($X_{\rm Na23}-X_{\rm O16}$) plane. The dashed,
dash-dotted, solid and dotted lines represent the evolution of
rapidly rotating stars with 1.0, 1.1, 1.2 and 1.3\,M$_{\rm \odot}$,
respectively. The evolutionary tracks of rapidly rotating stars
begin from the bottom right corner to top left corner. During the
evolution of these stars, they show an increase of sodium abundance
by $\sim$0.67\,dex and a decrease of oxygen abundance by
$\sim$0.74\,dex. Therefore, they can reach significant oxygen
depletion and sodium enhancement, although their variation ranges of
surface abundances are different and depend on their masses. Open
circles present the evolution of single stars that do not show
significant variations. } \label{fig2}
\end{figure}

\clearpage

\begin{figure}
\epsscale{0.9} \plotone{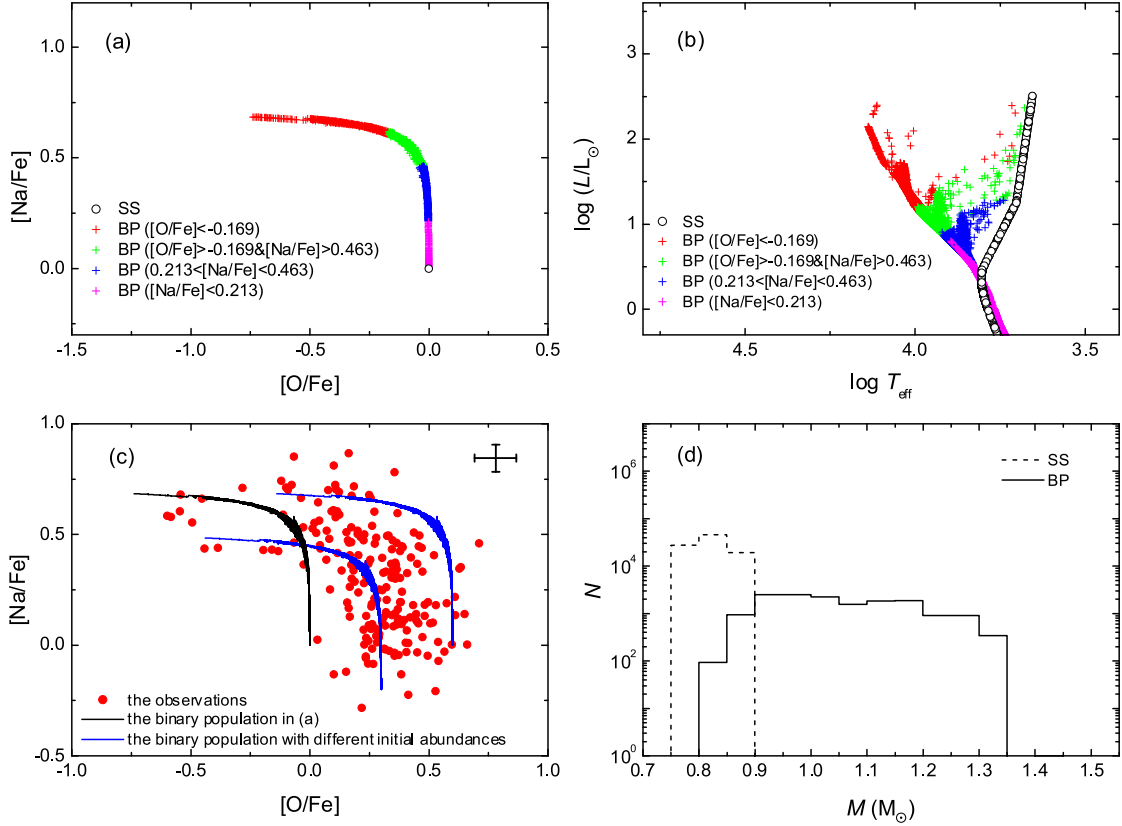} \caption{Distribution of single
stars (SS) and binary products (BP) in the simulation of the binary
population with binary interactions at 10\,Gyr. (a) the simulated
distribution in the surface abundances (Na-O) plane; (b) the
simulated HR diagram; (c) the observational abundances for 202 red
giants in 17 GCs \citep{Carretta2009}; (d) the simulated
distribution of mass. For panels (a) and (b), open circles
correspond to single stars, and pluses to binary products that are
classified into four group (red, green, blue, and magenta) according
to their surface mass fractions of Na and O. In panel (c), red
points show the observed red giants in 17 GCs given by Carretta et
al. (2009), and typical errors are indicated in this panel. Black
line presents the binary population in panel (a), for comparison.
Blue lines show two binary populations with different initial
abundances that are assumed to have the same variation ranges of
surface abundances of the binary population in panel (a). In panel
(d), the solid histograms and the dashed histograms are the
distributions of the masses for binary products and single stars
with log $L/L_{\rm \odot}
> 0$. \label{fig1}}
\end{figure}

\clearpage

\end{document}